\documentclass{KapProc}
\usepackage[xdvi]{graphicx}
\normallatexbib

\def\Red  {}
\def\Black{}
\def\Blue {}
\begin{document}
\articletitle[\Red Control of squeezed light pulse spectrum \Black]{\Red Control of
squeezed light pulse spectrum in the Kerr medium\\ with an inertial
nonlinearity\Black}
\author{A.S. Chirkin and F. Popescu}
\affil{Moscow State University,\\ 119899,\ Moscow,\
       Russia}
\email{chirkin@foton.ilc.msu.su\ \ \ florentin\_p@hotmail.com}
\begin{abstract}
\Blue The consistent quantum theory of self-phase modulation ({\bf SPM}) and
cross-phase modulation ({\bf XPM}) for ultrashort light pulses ({\bf USP}s) in
 medium with electronic Kerr-nonlinearity are developed. The approach makes
use of momentum operator of electrical field which takes account of the inertial
behaviour of the nonlinearity. The spectrum of quantum fluctuations of squeezed-quadrature
component as a function of response time of nonlinearity and values of nonlinear
phase shifts due to the {\bf SPM} and {\bf XPM} effects, is
investigated.\Black
\end{abstract}
\section{Introduction}
There are some methods to obtain USPs in a nonclassical state. One is the degenerate
three-frequency parametric amplification which has high phase sensitivity. Other is
SPM in a Kerr-nonlinear medium which in comparison with parametrical processes, does
not require ~phase-matching. This property is a real advantage for the pulsed
nonclassical state formation. The both participation of SPM and dispersion in the
nonlinear medium leads to the optical soliton formation \cite{AKH}. The quantum
theory of SPM and XPM for USPs developed during last $15$ years have difficulties
produced, in particular, by the multi-frequency structure of USPs and nonlinear kind
of their interaction.

The aim of our study is to develop the consistent quantum theory of nonlinear
propagation of USPs. An account is taken of the role of a response time of the
nonlinearity. The first attempt has been made in \cite{BLO}\ to give  quantum
analysis of SPM of USPs for the case of relaxation nonlinearity of medium, and has been
noted that the correct quantum theory should take account of additional noise sources
connected with nonlinear absorption. However, the theory developed in \cite{BLO}\
does not take account of noise sources, so that the commutation relation for
annihilation and creation photon operators is not fulfilled. This was carried out in
the developed in \cite{BOI} quantum theory of SPM, where noise sources are considered
as a fluctuation addition to the relaxation nonlinearity of medium. Therefore, the
results of \cite{BOI}\ are connected with the range of carrier frequencies of the
pulse, where the nonlinear absorption is important. Actually the authors \cite{BOI}\
have developed the quantum theory of USP propagation in a Raman active medium. In
\cite{BOI}\ the electronic Kerr nonlinearity with finite response time is modeled as
the Raman active medium. Note, if we deal with the USPs propagation, for example,
through fused-silica fibres only about $0.2$ of the Kerr effect is attributable to
the Raman oscillators and about $0.8$ of the Kerr effect is due to electronic motion
\cite{SHA}. The case when the USP carrier frequency is far enough from some
resonances (one- and two-photon and the Raman resonances), and therefore absorption
is absent, was investigated by us in \cite{POP1}\ where the finite response time is
considered in the interaction Hamiltonian, and the commutation relation is exactly
fulfilled.

In the present work the results of the consistent quantum theory of SPM and XPM of
USPs based on the use of the momentum operator (quantity of movement) are presented.
The developed approach can be used when the response time is much shorter than the
pulse duration and dispersion in nonlinear medium is neglected. There
are no restrictions on the pulse intensity in our theory.
\section{Quantum theory of SPM of USP{\scriptsize{s}}}
The traditional way to describe the SPM of USP is based on the interaction
Hamiltonian use, at that  usually we solve the time-evolution equation. The transition
to the spatial-evolution equation is realized substituting $t$ with $z/u$, where $z$
is the distance passed in medium and $u$ is the group velocity. This approach seems to be
good enough for single-mode radiation. If we deal with nonlinear propagation of USP
then both $t$ and $z$ are present in analytical description. Thus, we use the
momentum operator of pulse field related to the space-evolution \cite{TOR}.

We describe the SPM of USP using the momentum operator (cf. \cite{POP1})
\begin{equation}\label{hamilt}
\hat{G}_{spm}(z)=\frac{1}{2}\hbar\beta\int_{-\infty}^{\infty}\,dt
\int_{-\infty}^{t}H(t-t_{1})\,{\hat{\mathbf{N}}}\left[\hat{n}(t,z)\hat{n}(t_{1},z)\right]\,dt_1,
\end{equation}
where $\hat{n}(t,z)=\hat{A}^{+}(t,z)\hat{A}(t,z)$,\ \  $\hat{A}^{+}(t,z)$
($\hat{A}(t,z)$) is the Bose operator creating (annihilating) photons in a given
cross-section $z$ of the medium at a given time $t$, $\hbar$ is the Planck's constant
and $\hat{\mathbf{N}}$ is the operator of normal ordering. The coefficient $\beta$ is
defined by the Kerr nonlinearity of a medium at stationary conditions. $H(t)$ is the
function of the nonlinear response of a medium; $H(t)\ne0$  at $t\ge0$ and $H(t)=0$
at $t<0$. The expression under the first integral in (\ref{hamilt}) can be
interpreted as generalized force acting in a defined cross-section $z$, which at the
moment of time $t$ depends only on the previous ones ($t_{1}\le t$), i.e. it
satisfies the causality principle. The nonlinear response function of a medium should
be introduced as :
\begin{equation}\label{has}
H(t)=(1/\tau_r)e^{-t/\tau_r} \qquad (t\ge0),
\end{equation}
where $\tau_{r}$ is the response time of nonlinear medium, that we assume to be much
shorter than the pulse duration $\tau_{p}$. This nonlinearity takes place in absence
of the one- and two- photons absorption and Raman resonances \cite{AKH}.
Besides, as it will be shown below, also in this limit case the account of finite
nonlinear response time plays an important role. The space evolution equation for
$\hat{A}(t,z)$ follows from (see \cite{TOR})
\begin{equation}\label{evolution}
-i\hbar\frac{\partial\hat{A}(t,z)}{\partial
z}=\left[\hat{A}(t,z),\hat{G}_{spm}(z)\right],
\end{equation}
and making use of (\ref{hamilt}), we finally get
\begin{equation}\label{arc}
\frac{\partial\hat{A}(t,z)}{\partial\ z}-i\frac{1}{2}\beta
q\left[\hat{n}(t,z)\right]\hat{A}(t,z)=0,
\end{equation}
where
\begin{equation}\label{qu0}
q\left[\hat{n}(t,z)\right]=\int_{0}^{\infty}H(t_1)
\left[\hat{n}(t-t_{1},z)+\hat{n}(t+t_{1},z)\right]\,dt_{1}.
\end{equation}
Here $\hat{n}(t,z)=\hat{A}^{+}(t,z)\hat{A}(t,z)$ is the photon number ``density". The
Eq.\ (\ref{arc}) describes SPM in the moving coordinate frame: $z=z^{'}$,
$t=t^{'}-z/u$, where $t^{'}$ is the running time. According to (\ref{arc}) the
operator $\hat{n}(t,z)$ does not depend on $z$, i.e.
$\hat{n}(t,z)=\hat{n}(t,z=0)=\hat{n}_{0}(t)=\hat{A}^{+}_{0}(t)\hat{A}_{0}(t)$ ($z=0$
corresponds to the entrance into the medium). Taking account of this fact
$q[\hat{n}(t,z)]=q[\hat{n}_0(t)]$, and (\ref{qu0}) can be rewritten as:
\begin{equation}\label{qu1}
q\left[\hat{n}_{0}(t)\right]=\int_{-\infty}^{\infty}h(t_1)\hat{n}_{0}(t-t_{1})\,dt_{1},\qquad
(h(t)=H(|t|)).
\end{equation}
Taking account of (\ref{qu1}), we get the solution of (\ref{arc})
\begin{equation}\label{A}
\hat{A}(t,z)=\exp{\left\{i\gamma q\left[\hat{n}_{0}(t)\right]\right\}}\hat{A}_{0}(t).
\end{equation}
For hermitian conjugate operator of $\hat{A}(t,z)$ we have
\begin{equation}\label{A+}
\hat{A}^{+}(t,z)=\hat{A}^{+}_{0}(t)\exp{\left\{-i\gamma
q\left[\hat{n}_{0}(t)\right]\right\}}.
\end{equation}
In expressions (\ref{A}), (\ref{A+}) $\gamma=\beta z/2$. For time-independent
operator $\hat{n}_{0}$, Eqs.\ (\ref{A}), (\ref{A+}) give us the results for single
mode radiation \cite{AHM}. In the case of the instantaneous response of the
nonlinearity $H(t)=\delta(t)$ we get \cite{BLO}\
\begin{equation}\label{equations}
\hat{A}(t,z)=e^{i2\gamma\hat{n}_{0}(t)}\hat{A}_{0}(t), \qquad
\hat{A}^{+}(t,z)=\hat{A}^{+}_{0}(t)e^{-i2\gamma\hat{n}_{0}(t)} .
\end{equation}
At the input to the medium Bose operators satisfy the commutation relation
$[\hat{A}_{0}(t_{1}),\hat{A}^{+}_{0}(t_{2})]=\delta(t_{1}-t_{2})$. In a correct
quantum methodology the analogical commutation relation must also be satisfied in
nonlinear medium, i.e. for any $z$
\begin{equation}\label{delta}
[\hat{A}(t_{1},z),\hat{A}^{+}(t_{2},z)]=\delta(t_{1}-t_{2}).
\end{equation}
The calculation of average values of (\ref{equations}) and the normal ordering are
followed by the non-integrable singularities appearance (for instance, $\delta(t)$
appears in the exponent). These shortcomings are absent using (\ref{A}) and
(\ref{A+}). Besides, their use requires knowledge of an algebra of time-dependent
Bose operators. The later has been developed in \cite{BLO,POP2}. The following
permutation relations \cite{POP2}\
\begin{eqnarray}
\hat{A}_{0}(t_{1})e^{\hat{O}(t_{2})}\!\!&=&\!\!e^{\hat{O}(t_{2})+i\gamma
h(t_{2}-t_{1})} \hat{A}_{0}(t_{1}),\label{A0}\\
e^{\hat{O}(t_{2})}\hat{A}^{+}_{0}(t_{1})\!\!&=&\!\!\hat{A}^{+}_{0}(t_{1})
e^{\hat{O}(t_{2})+i\gamma h(t_{2}-t_{1})},\label{A0+}
\end{eqnarray}
{}(where $\hat{O}(t)=i\gamma q[\hat{n}_{0}(t)]$) and the theorem of normal ordering \cite{BLOW}
\begin{equation}\label{teo}
e^{\hat{O}(t)}=\hat{\mathbf{N}}\exp{\biggl\{\int_{-\infty}^{\infty}\left[e^{i\gamma\tilde{h}(\theta)}-1\right]
\hat{n}_{0}(t-\tau_{r}\theta)\,d\theta\biggl\}},
\end{equation}
are valid, where $\tilde{h}(\theta)=e^{-|\theta|}$, $\theta=t/\tau_{r}$. As
$[\hat{O}(t_{1}),\hat{O}(t_{2})]=0$, then
$e^{\hat{O}(t_{1})}e^{\hat{O}(t_{2})}=e^{\hat{O}(t_{1})+\hat{O}(t_{2})}$. Making use
of relations (\ref{A0}) and (\ref{A0+}), one can verify that the commutation relation
(\ref{delta}) is exactly fulfilled.

As we analyse the SPM of an initial coherent USP, the equation on eigenvalues
$\hat{A}_{0}(t)|\alpha(t)\rangle=\alpha(t)|\alpha(t)\rangle$ is satisfied by
$\hat{A}_{0}(t)$ (see \cite{AKH}), where $|\alpha(t)\rangle$ is the initial coherent
state, $\alpha(t)$ is the eigenvalue of $\hat{A}_{0}(t)$, and
$|\alpha(t)|^{2}=\bar{n}_{0}(t)$ is the average photon density. Particular attention
is given to the quantum fluctuations behaviour of $X$-quadrature;
$\hat{X}(t,z)=[\hat{A}(t,z)+\hat{A}^{+}(t,z)]/2$. The behaviour of another quadrature
component is shifted in phase with $\pi/2$. For correlation function (see
\cite{POP1})
\begin{equation}
R(t,\tau)=\langle\hat{X}(t,z)\hat{X}(t+\tau,z)\rangle-
\langle\hat{X}(t,z)\rangle\langle\hat{X}(t+\tau,z)\rangle,\label{corelfunction}
\end{equation}
where the brackets denote averaging over the initial coherent state of the pulse,
 making use of
(\ref{A0})-(\ref{teo}) we get
\begin{equation}\label{recorel}
R(t,\tau)=\frac{1}{4}\Bigl[\delta(\tau)-\psi(t)h(\tau)\sin{2\Phi(t)}
+\psi^{2}(t)g(\tau)\sin^{2}{\Phi(t)}\Bigl].
\end{equation}
In (\ref{recorel}) the following notations have been introduced:
$\Phi(t)=\psi(t)+\varphi(t)$, $\varphi(t)={arg}\,\,{\alpha(t)}$ - the phase of USP or
the heterodyne pulse's phase at balanced homodyne detection,
$\psi(t)=2\gamma\bar{n}_{0}(t)=\psi_{0}\rho^{2}(t)$ - the nonlinear phase shift
resulted from SPM, $\psi_{0}=\psi(0)=2\gamma\bar{n}_{0}$ - the maximum nonlinear
phase shift, $\rho(t)$ - the envelope of USP ($|\alpha(t)|=\alpha_{0}\rho(t)$,
$\rho(0)=1$), and
$g(\tau)=\int_{-\infty}^{\infty}\tilde{h}(\theta)\tilde{h}(\theta+\tau)\,d\theta$.
The derivation of (\ref{recorel}) took into consideration the fact that, in many
experimental situations nonlinear phase shift per photon $\gamma\ll1$. The
instantaneous spectral density of the quantum fluctuations of $X$-quadrature
component $S_{X}(\omega,t)=\int_{-\infty}^{\infty}R_{X}(t,\tau)e^{i\omega\tau}d\tau$,
according to (\ref{recorel}), takes the form
\begin{equation}\label{sio}
S_{X}(\Omega,t)=\frac{1}{4}\Bigl[1-2\psi(t)L(\Omega)\sin{2\Phi(t)}
+4\psi^{2}(t)L^{2}(\Omega)\sin^{2}{\Phi(t)}\Bigl],
\end{equation}
where $L(\Omega)=1/[1+\Omega^{2}]$, $\Omega=\omega\tau_{r}$. {}From (\ref{sio}) it
follows that the quantum fluctuations level bellow the short noise one depends on
the nonlinear phase shift $\psi(t)$ and $\Phi(t)$. At the initial phase
of the pulse chosen optimal for a frequency
$\Omega_{0}=\omega_{0}\tau_{r}$
\begin{equation}\label{optphase}
\varphi_{0}(t)=\frac{1}{2}\arctan{\left[\frac{1}{\psi(t)L(\Omega_{0})}\right]}-
\psi(t),
\end{equation}
the  spectral density (\ref{sio}) is
\begin{equation}\label{specreduce}
S_{X}(\Omega_{0},t)=\frac{1}{4}\Bigl\{\left[1+
\psi^{2}(t)L^{2}(\Omega_{0})\right]^{1/2}-\psi(t)L(\Omega_{0})\Bigl\}^{2}.
\end{equation}
{}From (\ref{specreduce}) it follows that the spectral density adiabatically changes
itself with the changing pulse's envelope and it is lower than $1/4$ which
corresponds to the coherent state of the pulse. {}From (\ref{specreduce}) we can also
see that at the optimal phase $\varphi_{0}(t)$ with the increasing of the nonlinear
phase shift  $\psi(t)$, the spectral density $S_{X}(\Omega_{0},t)$ monotonously
decreases. We point out that at $\tau_{r}=0$ from (\ref{sio}) the results for single
mode radiation can be obtained, at that the quantum fluctuation spectrum's level does
not depend on the frequency. According to (\ref{specreduce}), increasing $\tau_{r}$
(increasing $\Omega_{0}$) at fixed frequency $\omega_{0}$, the quadrature
fluctuations level increases.
\section{Quantum theory of XPM of USP\scriptsize{s}}
\begin{figure}[ht]
\vskip2pt
\centering
\includegraphics[width=7.5cm,height=5.5cm]{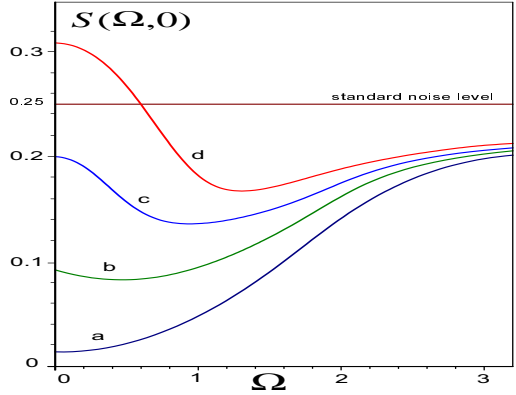}
\caption{\protect{The spectrum of quantum fluctuations of the investigated light
pulse at SPM and XPM as a function of reduced frequency $\Omega=\omega \tau_r$ for
different relations between average photon numbers of pulses $\bar{n_{0,2}}/\bar
{n_{0,1}}$: 0 (a), 3 (b), 5 (c), 8 (d). The graphics are depicted for the case
$\gamma_1=\gamma_2=2 \tilde {\gamma}$, and nonlinear phase shift $\psi_{0,1}(0)=2$ at
$t=0$. At SPM only (curve (a)), the maximum squeezing is realized for
$\Omega=0$.}}\label{fig1}
\end{figure}

We analyse now two-pulse propagation in an inertial nonlinear medium.  We will
consider that pulses have orthogonal polarizations or/and different frequencies.
Then, besides SPM of each pulse, the XPM effect takes place. Here we assume that
the parametrical interaction of pulses can be neglected. In this case, the analysed
process can be depicted making use of the following momentum operator \cite{PCross}.
\begin{equation}\label{spm}
\hat{G}(z)=\sum_{j=1}^{2}\hat{G}^{(j)}_{spm}(z)+\hat{G}^{(1,2)}_{xpm}(z).
\end{equation}
Here $\hat{G}^{(j)}_{spm}(z)$ is due to the SPM of $j$-pulse ($j=1,2$) (see
\ref{hamilt}) and operator $\hat{G}^{(1,2)}_{xpm}(z)$ is connected with the XPM of
pulses
\begin{equation}\label{hcrossphase}
\hat{G}^{(1,2)}_{xpm}(z)\!=\hbar\widetilde{\beta}\int_{-\infty}^{\infty}\!dt
\int_{-\infty}^{t}\!\!H(t\smash{-}t_{1})[\hat{n}_{1}(t,z)\hat{n}_{2}(t_{1},z)
+\hat{n}_{2}(t,z)\hat{n}_{1}(t_{1},z)]dt_{1}.
\end{equation}
In (\ref{hcrossphase}) $\hat{n}_{j}(t,z)$ is the photon number operator of $j$-pulse
and coefficients $\beta_{j}$, $\widetilde{\beta}$ are responsible for the SPM and XPM
respectively. As stated earlier, the operator $\hat{n}_{j}(t,z)$ commutes with
$\hat{G}(z)$. In consequence $\hat{n}_{j}(t,z)=\hat{n}_{j}(t,z=0)=\hat{n}_{0,j}(t)$,
which means that the photons statistics of each pulse remains unchanged in the
nonlinear medium. In accordance with (\ref{evolution}), the evolution of
$\hat{A}_{1}(t,z)$ for the first pulse is given by (cf. (\ref{arc}))
\begin{equation}\label{evol}
\frac{\partial\hat{A}_{1}(t,z)}{\partial z}-i\left\{\frac{1}{2}
\beta_{1}q\left[\hat{n}_{0,1}(t)\right]+\widetilde{\beta}q\left[\hat{n}_{0,2}(t)\right]\right\}\hat{A}_{1}(t,z)=0.
\end{equation}
One can get for $\hat{A}_{2}(t,z)$ the similar equation changing the index $1\mapsto
2$ in (\ref{evol}). The functions $q[\hat{n}_{0,j}(t)]$ is similar to (\ref{qu1}). We
remark that (\ref{arc}), (\ref{evol}), are written in the moving frame. The solution
of (\ref{evol}) is
\begin{equation}\label{Across}
\hat{A}_{1}(t,z)=e^{i\gamma_{1}q[\hat{n}_{0,1}(t)]+
i\widetilde{\gamma}q[\hat{n}_{0,2}(t)]}\hat{A}_{0,1}(t),
\end{equation}
where $\gamma_{1}=\beta_{1}z/2$, $\widetilde{\gamma}=\widetilde{\beta}z$. We define
the correlation function of the investigated light pulse as (\ref{corelfunction}).
For the spectra of quantum fluctuations of $X_{1}$- quadrature we get
\begin{eqnarray}
S_{X_{1}}(\Omega,t)=\frac{1}{4}\Bigl\{1\!\!\!\!&-&\!\!\!\!2\psi_{1}(t)L(\Omega)\sin{2\Phi_{1,2}(t)}\nonumber\\
&+&\!\!\!4[\psi^{2}_{1}(t)\smash{+}\widetilde{\psi}_{1}(t)\widetilde{\psi}_{2}(t)]
L^{2}(\Omega)\sin^{2}{\Phi_{1,2}(t)}\Bigl\}.\label{spectrale}
\end{eqnarray}
Here we denoted: $\Phi_{1,2}(t)=\Phi_{1}(t)+\widetilde{\psi}_{2}(t)$,
$\Phi_{1}(t)=\psi_{1}(t)+\varphi_1(t)$, $\psi_{1}(t)=2\gamma_{1}\bar{n}_{0,1}(t)$,
$\widetilde{\psi}_{j}(t)=2\widetilde{\gamma}\bar{n}_{0,j}(t)$ ($j=1,2$). As already
mentioned, the phase $\Phi_{1}(t)$ is connected with own parameters of the
investigated pulse, and the phase $\widetilde{\psi}_{1}(t)$ with the XPM. {}From
comparison of (\ref{spectrale}) and (\ref{sio}) one can see that the XPM adds new
terms in the multiplier and in the phase in (\ref{spectrale}). This circumstance allows us to
control the fluctuations spectrum of the investigated pulse. At the initial phase
\begin{equation}
\varphi_{0,1}(t)=\frac{1}{2}\arctan{\biggl[\frac{\psi_{1}(t)}
{\psi^{*}_{1,2}(t)L(\Omega_{0})}\biggl]}
\smash{-}\psi_{1}(t)\smash{-}\widetilde{\psi}_{2}(t),
\end{equation}
chosen optimal for frequency $\Omega_{0}=\omega_{0}\tau_{r}$, the spectral density
(\ref{spectrale}) is
\begin{equation}
S_{X_{1}}(\Omega_{0},t)=\frac{1}{4}\Bigl\{1-2L(\Omega_{0})
\left[\psi^{2}_{1}(t)\smash{+}\psi^{{*}^{2}}_{1,2}(t)L^{2}(\Omega_{0})\right]^{1/2}\!
-\!\psi^{*}_{1,2}(t)L(\Omega_{0})\Bigl\},
\end{equation}
where
$\psi^{{*}}_{1,2}(t)=\psi^{2}_{1}(t)+\widetilde{\psi}_{1}(t)\widetilde{\psi}_{2}(t)$.
The influence of the second pulse to the spectrum of the first one is depicted in
{\it Figure \ref{fig1}}.
Increasing the photon number (intensity) of the second pulse, their quantum
fluctuations can substantially increase the level of this ones for the investigated
pulse. Changing the intensity of the second pulse, one can control the spectrum of
quantum fluctuations of the the investigated pulse's quadrature.
\section{Conclusions}
The main result of our work is represented by the development of the consistent
quantum theory of SPM and XPM of USPs in non-absorption nonlinear media, when the
carrier frequencies of the pulse is far enough from any resonances. However, the
consideration of the finite response time of nonlinearity is necessary to get correct
solutions ((\ref{A}), (\ref{A+}), (\ref{Across})). Our results are valid for response
time of nonlinearity much shorter than duration of pulses  but for unspecified
intensities
 of USPs. In
some sense, our results are complementary with the ones presented in \cite{BOI} which
treats the situation when the Raman resonance is important. The quantum theory
of XPM of USPs in the developed approach here is presented for the first time. We
have shown that the form and the level of the fluctuation spectrum can
be controlled with the change of the pulse's phase and
 the another pulse's intensity in the presence of XPM. This fact can be used
for quantum non-demolition measurements of parameters of the pulse \cite{WAL}.
\begin{acknowledgments}
\Blue The authors are grateful to K.N. Drabovich (MSU, Moscow) for useful discussions and to
S. Codoban (JINR, Dubna) for rendered help. This work has been partial supported by
Programme ``Fundamental Metrology''\Black.
\end{acknowledgments}


\begin{chapthebibliography}{1}
\bibitem{AKH} S.A. Akhmanov, V.A. Vysloukh, and A.S. Chirkin, {\it Optics of
              Femtosecond Laser Pulses}, N.Y.: AIP, 1992.
\bibitem{BLO} K.J. Blow, R. Loudon, and S.J.D. Phoenix,  J. Opt. Soc. Am. B, 8:1750, 1991.
\bibitem{BOI} L. Boivin, F.X. K\"{a}rtner, and H.A. Haus, Phys. Rev. Lett., 73:240,
              1994;  L. Boivin, Phys. Rev. A., 52:754, 1994.
\bibitem{SHA}  L.G. Joneckis, and J.H. Shapiro, J. Opt. Soc. Am. B, 10:1102, 1993.
\bibitem{POP1} F. Popescu, A.S. Chirkin, Pis'ma Zh. Eksp. Teor. Fiz., 69:481, 1999.
               [JETP Lett. 69:516, 1999].
\bibitem{TOR} M. Toren, and Y. Ben-Aryeh, Quantum Opt., 9:425, 1994.
\bibitem{AHM} S.A. Akhmanov, A.V. Belinskii, and A.S. Chirkin, in {\it
              New Physical Principles of Optical Information Processing} (in Russian).
              Eds: S.A. Akhmanov, M.A. Vorontsov, Nauka, Moscow, p. 83, 1990.
\bibitem{POP2} F. Popescu, and A.S. Chirkin, Phys. Rev. A. (to be published) (see.
               LANL E-print: quant-ph/0003028), 2000.
\bibitem{BLOW} K.J. Blow, R. Loudon, S.J.D. Phoenix, and T.J. Shepherd, Phys. Rev. A.,
               42:4102, 1990.
\bibitem{PCross} F. Popescu, and A.S. Chirkin, LANL E-print: quant-ph/0004049 (to be
                 published), 2000.
\bibitem{WAL}  D.F. Walls, and G.J. Milburn, {\it Quantum Optics}, Springer, Berlin, 1995.
\end{chapthebibliography}
\end{document}